\documentclass[12pt]{article}

\pdfoutput=1

\usepackage{amssymb}
\usepackage{amsmath}
\usepackage[top=80pt,bottom=85pt,left=85pt,right=85pt]{geometry}
\usepackage{setspace}
\usepackage{sectsty}
\usepackage{braket}
\usepackage{graphicx,color,subfigure}
\usepackage{xcolor}
\usepackage{mathalfa}
\usepackage{cancel}
\usepackage[debug,pageanchor=false]{hyperref}
\definecolor{link}{rgb}{.8,.15,.1}
\usepackage{mdframed}
\usepackage{soul}
\usepackage{multirow}
\usepackage[section]{placeins}
\hypersetup{colorlinks=true,linkcolor=link,citecolor=link,urlcolor=link,linktocpage}

\newcommand{\LL}{\mathcal{L}}
 
\newcommand{\rr}{\mathbb{R}}
\newcommand{\zz}{\mathbb{Z}}
\newcommand{\R}{\mathrm{Re}}

\newcommand{\OSp}{$\mathfrak{osp}$}
\newcommand{\SU}{$\mathfrak{su}$}
\newcommand{\SO}{$\mathfrak{so}$}

\makeatletter
\@addtoreset{equation}{section}
\makeatother

\begin{document}

	\begin{titlepage}

	\begin{center}

	\vskip .5in 
	\noindent

	{\Large \bf{Spin-2 spectrum of six-dimensional field theories}}

	\bigskip\medskip

	Achilleas Passias and Alessandro Tomasiello\\

	\bigskip\medskip
	{\small 
Dipartimento di Fisica, Universit\`a di Milano--Bicocca, \\ Piazza della Scienza 3, I-20126 Milano, Italy \\ and \\ INFN, sezione di Milano--Bicocca
	}

	\vskip .5cm 
	{\small \tt achilleas.passias, alessandro.tomasiello@unimib.it}
	\vskip .9cm 
	     	{\bf Abstract }
	\vskip .1in
	\end{center}

	\noindent
	 We analyze the mass spectrum of spin-2 excitations around the gravity duals of ``linear quiver'' supersymmetric conformal field theories (SCFT's) in six dimensions. We show that for the entire family of gravity solutions it satisfies a bound which corresponds to a unitarity bound for the scaling dimension of the dual field theory operators. We determine the masses of excitations which belong to short multiplets,  and for certain gravity solutions we obtain the Kaluza--Klein modes and the corresponding mass spectrum, fully and explicitly. Finally, we discuss an intuitive picture of the dual operators in terms of the effective descriptions of the SCFT's.

	\noindent

	\vfill
	\eject

	\end{titlepage}
\tableofcontents

\section{Introduction}

Interacting quantum field theories in higher than four dimensions are hard to define, non-renormalizability being a typical obstacle. Given how little we know about field theory in higher dimensions, any example which is even partially amenable to quantitative analysis is welcome. 

In this paper, we focus on a class of six-dimensional conformal field theories (CFT's) with ${\cal N}=(1,0)$ supersymmetry. They were originally introduced  \cite{hanany-zaffaroni-6d,brunner-karch,blum-intriligator,intriligator-6d,intriligator-6d-II} in various string theory setups, one of which is a network of NS5, D6 and D8-branes. Although the CFT's  do not have a Lagrangian description, they have a ``Coulomb'' or tensor branch along which an effective Lagrangian can be written down. This consists of several vector, tensor and hyper multiplets arranged in a linear chain or ``quiver''. Recently it has been argued \cite{gaiotto-t-6d} that their holographic duals belong to a class of AdS$_7$ solutions of massive type IIA supergravity, found numerically in \cite{Apruzzi:2013yva} and analytically in \cite{Apruzzi:2015wna}. A check of this identification was later performed in \cite{cremonesi-t}, where the Weyl anomaly was computed on both sides, finding agreement.\footnote{Section 2 of that paper gives a review of the solutions and of the holographic dictionary.}

Having identified this holographic duality, we can use it to compute certain features of the field theory by performing gravity computations. For instance, the Kaluza--Klein (KK) spectrum of a gravity solution is holographically dual to the operator spectrum of the field theory. 

Unfortunately, the full KK reduction is usually quite involved. Schematically, the task can be divided in two stages. In the first stage, one writes the linearized equations of motion for perturbations around the gravity solution AdS$_d \times M_{D-d}$ and reorganizes them as equations of motion for free fields in AdS$_d$, with masses given by the eigenvalues of certain ``mass operators''  which are differential operators on $M_{D-d}$. At this stage one can often be quite general, using only generic features of the class of solutions under consideration, and non-committing to a precise space $M_{D-d}$. In the second stage one computes the spectrum of the mass operators on a specific $M_{D-d}$. For a review, see for example \cite{duff-nilsson-pope} for $D=11$, $d=4$.\footnote{The first stage is reviewed there for the class of Freund--Rubin solutions, and culminates with Table 5. The second stage is reviewed more specifically for $M_7=S^7$. Notable examples of KK reductions for $D=10$, $d=5$ include \cite{kim-romans-vannieuwenhuizen,Ceresole:1999ht,Ceresole:1999zs}.} While this program has been completed for the AdS$_7\times S^4$ solution of $D=11$ supergravity, which is dual to the $(2,0)$ theory on the M5-brane worldvolume, for the AdS$_7\times M_3$ solutions of \cite{Apruzzi:2013yva,Apruzzi:2015wna} it would have to be started from scratch, and it is made more complicated by the presence of a non-trivial warping function, fluxes not of the Freund-Rubin type, and by possible contributions from internal D-branes. 

In this paper we consider what is arguably the simplest part of this program: the spectrum of massive spin-2 particles or gravitons in AdS$_7$. For these particular fields, the mass operator on the internal space $M_3$ is a scalar Laplacian, modified in a certain way by the warping function (which encodes the dependence of the AdS radius on $M_3$). This fact has already been used in several cases, to infer properties of the corresponding operator spectrum, without performing the full KK reduction; see for example \cite{Klebanov:2009kp, Ahn:2009et, Ahn:2009bq, Bachas:2011xa, Richard:2014qsa, pang-rong} for AdS$_4$ solutions. In particular, \cite{Bachas:2011xa} showed in general the above statement about the spin-2 mass operator, drawing on earlier results by \cite{Csaki:2000fc}, and their arguments can be applied also to our case. 

The states we find are dual to spin-2 operators, but supersymmetry relates them to various other operators, among which scalars. The resulting multiplets have a rather intricate structure, which we have found by reducing the multiplets of the $(2,0)$ superconformal algebra which occur in the AdS$_7\times S^4$ solution and have been explicitly constructed (see appendix \ref{multiplets}). As predicted by general representation theory for the $(1,0)$ superconformal algebra \cite{Minwalla:1997ka,dobrev}, these multiplets are shortened when the scaling dimension and the representation of the R-symmetry \SU(2)$_{\rm R}$ are related.\footnote{Besides the ones we discuss in appendix \ref{multiplets}, there are other multiplets that contain spin-2 operators; see for example \cite[Table 31]{cordova-dumitrescu-intriligator-def} for a list of short ones. However, they typically also contain fields with spin higher than two (namely, fields with Dynkin labels $D_1+D_2+D_3>2$), which would never appear in a KK supergravity reduction. We thank C.~Cordova for checking this for us.\label{foot:clay}} 

One of the results of this paper is a bound on the mass spectrum for the entire class of gravity solutions. The dimension of the corresponding operator in  a generic theory in the linear quiver class is then bounded by its  SU(2)$_{\rm R}$-charge, a bound that is in fact a unitarity bound. Moreover, there is always one short spin-2 multiplet for each SU(2)$_{\rm R}$ representation, labeled by its ``spin'' $\ell$. The spin-2 operator in this multiplet has scaling dimension $\Delta=4 \ell + 6$. In the effective theory, an operator of the correct dimension and SU(2)$_{\rm R}$ representation would be of the form ${\rm Tr}(h^{(I_1}h^{I_2\,\dagger}\ldots h^{I_{2\ell-1}} h^{I_{2\ell})\,\dagger}) T_{\alpha \beta}$, with $h^I$ being scalars in hypermultiplets, $I_i=1,2$ SU(2)$_{\rm R}$ doublet indices, and $T_{\alpha \beta}$ the stress-energy tensor. The superconformal primary in the multiplet is then a scalar of scaling dimension 
\begin{equation}
	\Delta_{\rm pr}= 4 \ell +4\  .
\end{equation}
This applies to all linear quiver theories.

For certain gravity solutions --- dual to very simple linear quiver theories --- we were able to find explicitly the complete spectrum, for both short and long multiplets. One solution is the dimensional reduction of an orbifold of the AdS$_7\times S^4$ solution of eleven-dimensional supergravity. In this case there is a second short multiplet with $\Delta_{\rm pr} = 4\ell+6$. Another one is a solution with Romans mass and one stack of D6-branes. For more general solutions, finding the complete spectrum (including long multiplets) is harder, but we demonstrate in one example with one stack of D8-branes how to set up the problem, and obtain some numerical results. Finally in an appendix we consider the non-supersymmetric solutions presented rather implicitly in \cite{prt} and recently interpreted holographically in \cite{apruzzi-dibitetto-tizzano}. These are in one-to-one correspondence with the supersymmetric ones, and we show that their spin-2 mass spectrum is easily related to those.

There are multiplets that we have not been able to access in our analysis, because they don't contain any spin-2 operators; these are the vector and gravitino multiplets in Tables \ref{vector} and \ref{gravitino}, appendix \ref{multiplets}. Obtaining the spectrum of these additional multiplets requires performing the full Kaluza--Klein reduction. As noted earlier, such a reduction is not a simple modification of a previous one. In particular, it should take into account the D6 and D8-branes, which are present in all the AdS$_7$ solutions. If one were to obtain the multiplets of all spins, and especially the short ones, one would be able to compute an index (as for example done in \cite{bhattacharya-bhattacharyya-minwalla-raju} for the $(2,0)$ theory) and perhaps compare it with a field theory computation. 

The rest of the paper is organized as follows: after a quick review of the AdS$_7$ solutions in section \ref{sec:ads7}, we describe the relevant mass operator in section \ref{sec:graviton}. We compute its spectrum for certain solutions in section \ref{sec:ex}, and describe the field theory implications in section \ref{sec:ft}. In appendix \ref{multiplets} we summarize the \OSp$(6,2|1)$ multiplets of supergravity fields. In appendix \ref{non-susy} we consider the non-supersymmetric solutions. Appendix \ref{hyperJac} contains some technical details.

\section{\texorpdfstring{The AdS$_7$ solutions}{The AdS(7) solutions}}
\label{sec:ads7}

In this section we review the AdS$_7$ solutions of type IIA supergravity discovered first numerically in \cite{Apruzzi:2013yva} and presented later analytically in \cite{Apruzzi:2015wna}.

The internal manifold $M_3$ of the solutions is topologically a three-sphere $S^3$, whose metric is a warped product of an interval $I$ and a two-sphere $S^2$, with the two-sphere shrinking at the endpoints of the interval. The solutions are characterized by a single function $\beta$ of the coordinate $y$ which parametrizes $I$. This function governs the way the $S^2$ shrinks at the endpoints of the interval, thus determining if the geometry there is regular or singular, and the type of the singularity. In particular there are three cases. Let $y_0$ be an endpoint of $I$; then (i) if $\beta = \beta_1 (y-y_0) + O(y-y_0)^2$  near $y_0$, the geometry of $M_3$ approximates that of $\mathbb{R}^3$ (regular endpoint) (ii) if $\beta = \beta_2(y-y_0)^2 + O(y-y_0)^3$ the geometry of $M_3$ approximates the geometry of the transverse space of  a stack of (anti-)D6-branes near the singularity (iii) if $\beta = \beta_0 + \beta_{1/2}(y-y_0)^{1/2} + O(y-y_0)$ the geometry of $M_3$ approximates the geometry of the transverse space  of  an O6-plane near the singularity.
The metric of the solutions, in string frame, reads\footnote{A prime here denotes differentiation with respect to $y$.}
\begin{equation}\label{metric}
ds^2_{10} = e^{2A}\left( ds^2_{\mathrm{AdS}_7} - \frac{1}{16} \frac{\beta'}{y\beta} \, dy^2 + \frac{1}{4}\frac{\beta}{4 \beta-y \beta'} ds^2_{S^2} \right) \ ,
\qquad
e^{2A}= \frac49 \left(-\frac{\beta'}{y}\right)^{1/2} \ .
\end{equation}
The anti-de Sitter metric is of radius one. The dilaton is given by
\begin{equation}
e^{2\phi} = \frac{1}{144}\frac{(-\beta'/y)^{5/2}}{4 \beta - y \beta'} \ .
\end{equation}
The active fluxes are the NS-NS three-form field strength $H$ and the R-R two-form field strength $F_2$:
\begin{subequations}
\begin{align}
	F_2 &= \frac{y\beta^{1/2}}{\beta'}\left(4-\frac{F_0}{18y}\frac{(\beta')^2}{4 \beta - y \beta'} \right){\rm vol}_{S^2} \ ,\\
	H &= -\left(-\frac y{\beta'}\right)^{1/4}\left(9+\frac{F_0}{12y}\frac{(\beta')^2}{4 \beta-y \beta'} \right){\rm vol}_{M_3} \ .
\end{align}
\end{subequations}

For zero Romans mass $F_0$, 
\begin{equation}\label{zeroF0}
\beta = \left(\frac{2}{k}\right)^2(y^2-y^2_0)^2\ ,
\end{equation}
where $k$ is an integer; this solution is the dimensional reduction of the AdS$_7 \times S^4/\mathbb{Z}_k$ solution of eleven-dimensional supergravity \cite{Apruzzi:2013yva}. At one endpoint of $I$ or pole of $M_3$ there is stack of $k$ D6-branes, and on the other one a stack of $k$ anti-D6-branes. We will encounter this solution in section \ref{zeroRomansmass}.

For non-zero Romans mass, $\beta$ depends on two parameters (apart from the Romans mass), one of which determines its behavior near the endpoints. For a certain range of this parameter there are solutions with an O6-plane singularity at one endpoint and a D6-brane singularity at the other. For another range, there are solutions with D6-brane singularities at both endpoints. These two families of solutions meet at a single value of the parameter, where the solution is regular at one pole and has a D6-brane singularity at the other pole. We will refer to this solution as $\mathbb{R}^3$--D6, the name referring to the approximate geometry near each of the two endpoints of $I$.

The general solution with $F_0\neq 0$ can be found in \cite{Apruzzi:2015wna}. For the particular case of the $\mathbb{R}^3$--D6 solution,
\begin{equation}\label{R3D6}
\beta = \frac{8}{F_0}(y-y_0)(y+2y_0)^2 \ .
\end{equation}
The parameter $y_0$ and the Romans mass $F_0$ have opposite signs. Without loss of generality $y_0$ is fixed to be negative and $F_0$ positive. The coordinate $y \in [y_0,-2y_0]$. At
$y=y_0$, the $S^2$ in \eqref{metric} shrinks in a regular way; at $y=-2y_0$ there is a singularity of the D6-brane type. The $\mathbb{R}^3$--D6 solution will be the subject of section \ref{sub:R3D6}.

The families of solutions described above can be expanded by introducing D8-brane sources. These have the effect of changing $F_0$ as they are crossed; in each region between two D8-branes, the parameters of $\beta$ have different values. The positions of the D8-branes are fixed by supersymmetry. In general, these solutions are obtained by gluing several copies of the general massive solution, and of (\ref{zeroF0}) in regions where $F_0=0$, if they are present. 
A simple example is a solution where there is only one stack of D8-branes \cite{Apruzzi:2015wna}. The Romans mass jumps from a value $F_0$ to a value $F_0'=F_0(1-N/\mu)$, where $N =\frac1{4\pi^2}\int H$ is the NS-NS flux integer, and $\mu$ is a second integer parameterizing the D6-brane charge of the D8-branes. The function $\beta$ of the solution is simply (\ref{R3D6}) on one side of the D8-brane stack, and the same expression on the other side with $(F_0\to F'_0, y_0\to y'_0)$. The parameters $y_0$, $y_0'$, and the position $y_{\rm D8}$ of the D8-brane stack are given by 
\begin{equation}\label{eq:1D8}
	y_0 = -\frac32 F_0 \pi^2 (N^2-\mu^2) \ ,\qquad y_0'=\frac32 F_0 \pi^2 (N-\mu)(2 N- \mu)
	\ ,\qquad y_{\rm D8}= 3 F_0 \pi^2 (N-2\mu)(N-\mu)\ .
\end{equation}
The analytic expressions for more complicated solutions with D8-branes can be found in complete generality in \cite{cremonesi-t}.

\section{The graviton spectrum}
\label{sec:graviton}

In this section we will perform the Kaluza--Klein expansion of anti-de Sitter gravitons. We will describe the relevant mass operator in section \ref{sub:mass}, and the space of functions it should act on in sections \ref{sub:bound} and \ref{sub:bc}.

\subsection{The mass operator} 
\label{sub:mass}

We are interested in perturbations $\delta \bar{g}_{\mu\nu}$ of the anti-de Sitter metric\footnote{We add a bar to distinguish from the ten-dimensional metric $g_{\mu\nu} = e^{2A}  \bar{g}_{\mu\nu}$.}
\begin{equation}
\bar{g}_{\mu\nu} \to \bar{g}_{\mu\nu} + \delta \bar{g}_{\mu\nu}\ ,
\end{equation}
of the factorised form
\begin{equation}\label{factorA}
\delta \bar{g}_{\mu\nu}(\bar{x},x) = h_{\mu\nu}(\bar{x}) Y(x) \ .
\end{equation}
Here $\bar{x}$ denotes collectively the coordinates of the anti-de Sitter spacetime, and $x$ the coordinates of the internal manifold; $h_{\mu\nu}$ is transverse and traceless 
\begin{equation}\label{eq:hY}
\bar{\nabla}^\mu h_{\mu\nu} = \bar{g}^{\mu\nu} h_{\mu\nu} = 0 \ ,
\end{equation}
and satisfies
\begin{equation}
\bar{\nabla}^\rho \bar{\nabla}_\rho h_{\mu\nu} + (2-M^2) h_{\mu\nu} = 0 \ .
\end{equation}
It describes a massive graviton or spin-2 particle propagating in AdS --- see for example \cite{Polishchuk:1999nh}.

We draw upon the result of Bachas and Estes \cite{Bachas:2011xa}, who showed that the linearized Einstein equations for such perturbations of any flux compactification whose geometry contains a factor with maximal symmetry ---  i.e. anti-de Sitter, Minkowski or de Sitter --- amount to a differential equation for $Y$:\footnote{Their result was obtained for compactifications to four dimensions but can easily be extended to any dimensions.}
\begin{equation}\label{Yeigen}
\mathcal{L} Y = - M^2 Y \ ,
\end{equation}
where $\mathcal{L}$ is second order differential operator (see below). The central point of the derivation of \cite{Bachas:2011xa} is that $\delta\bar{g}_{\mu\nu}$ decouples from all other perturbations of the background solution; this is reflected in the variation of the stress-energy tensor which reads  $\delta T_{\mu\nu} \propto {\rm tr}( T) \, \delta \bar{g}_{\mu\nu}$, with ${\rm tr}( T)$ being the trace of the stress-energy tensor of the background solution. Notice that this is also true for local sources of the stress-energy tensor, such as D-branes or O-planes.

The result of \cite{Bachas:2011xa} was obtained in the Einstein frame to which we henceforth switch after a rescaling
\begin{equation}
g_{{\rm Einstein}} = e^{-\phi/2} g_{{\rm string}}
\end{equation}
of the string frame metric presented in the previous section.

The analysis of the mass spectrum of $h_{\mu\nu}$ thus becomes an eigenvalue problem \eqref{Yeigen} 
where $\mathcal{L}$ is a modified Laplacian on the internal manifold $M_3$
\begin{equation}\label{Lope}
\mathcal{L} \equiv \frac{e^{-5A+2\phi}}{\sqrt{\hat{g}}} \partial_{m} \left(e^{7A-2\phi}\sqrt{\hat{g}} \hat{g}^{mn} \partial_n\right),
\end{equation}
$\hat{g}$ being the metric on $M_3$.

We can reduce the eigenvalue problem to an ordinary differential equation (ODE) by expanding $Y$ in terms of the $S^2$ spherical harmonics:
\begin{equation}\label{factorY}
Y = \sum_{\ell=0}^\infty \sum_{m=-\ell}^\ell \beta^{\ell/2} f_{\ell,m}(y) Y_\ell^m.
\end{equation}
The factor $\beta^{\ell/2}$ is included as it simplifies the resulting ODE: dropping the labels $\ell$ and $m$ from $f$, the latter satisfies
\begin{equation}\label{ODE}
S f = - \lambda w(y) f \ , 
\end{equation}
where
\begin{align}\label{eq:S}
S \equiv \frac{d}{dy}\left(p(y)\frac{d}{dy}\right) , \qquad
p(y) \equiv -\frac{y \beta^{\ell + 3/2}}{\beta'} , \qquad
w(y) \equiv \beta^{\ell + 1/2} ,
\end{align}
and
\begin{equation}
	\lambda \equiv \frac{1}{16}[M^2-4\ell(4\ell+6)]\ .
\end{equation}
The variable $y$ takes values in the closed interval $I \equiv [y_-,y_+]$. This is called a \emph{singular} Sturm--Liouville problem: $p$ vanishes at the endpoints of $I$ for all $\beta$.


\subsection{Normalizability, and a bound on the spectrum}
\label{sub:bound}

We should now define a space of admissible solutions to \eqref{ODE}. Here we will give a definition that is naturally suggested by the theory of Sturm--Liouville problems; in the next subsection we will check that it agrees with physics intuition. 

Let us first define the weighted inner product
\begin{equation}\label{innerp}
(f_1,f_2)_{w} \equiv \int_I  f_1(y) f_2(y) \, w(y) dy\ .
\end{equation}
Suppose now $f_1$ and $f_2$ satisfy (\ref{ODE}), with different ``eigenvalues'' $\lambda_1$, $\lambda_2$. Then, integrating by parts we see that
\begin{align}\label{eq:orth}
	(\lambda_2-\lambda_1)(f_1,f_2)_w= \int_I (S f_1)f_2-f_1 (S f_2) =\left[p(f'_1 f_2-f_1 f_2')\right] \big|^{y_+}_{y_-}\ .
\end{align}
As noted earlier, the function $p$ is zero at the two endpoints $y_-$ and $y_+$,  so that if we pick boundary conditions such that 
\begin{equation}\label{eq:bc}
	f,f' \ \text{remain finite at} \ y_\pm  \ ,
\end{equation}
the right-hand side of (\ref{eq:orth}) vanishes, and hence two eigenfunctions with different eigenvalues are orthogonal.\footnote{Formally, $S$ is self-adjoint with respect to the ordinary inner product $(f_1,f_2)=\int_I f_1 f_2$. Most versions of the spectral theorem require a stronger definition of self-adjointness: the domain of the adjoint of $S$ would be defined via Riesz' theorem as $D(S^\dagger)=\{f | (Sf,\cdot) \ \text{is continuous} \}$, and one would then require $D(S^\dagger)=D(S)$ as well as $S=S^\dagger$. This is subtle to obtain for singular problems such as ours, but we will not need these details in what follows.} We will thus work in the space $L^2_w(I)$ of square-integrable functions with respect to the norm defined by this inner product (or more precisely in the corresponding Sobolev space where the norms of the first two derivatives are also finite).

Another way to motivate \eqref{innerp} is to start from the  Klein--Gordon inner product for ten-dimensional scalars; this is the approach of  \cite{Bachas:2011xa}. Employing a factorization similar to \eqref{factorA}, \eqref{factorY}, the inner product induced on $f$ is \eqref{innerp}.

We will now derive a key bound on the mas spectrum $M^2$. Given $f$ an eigenfunction of $S$, we have
\begin{equation}
\lambda (f,f)_w = \int_I (-S f) f dy = \int_I p f'^2 dy - \left[p f' f\right]\big|_{y_-}^{y_+} \ .
\end{equation}
Just like for (\ref{eq:orth}), the $\left[p f' f\right] \big|_{y_-}^{y_+}$ term vanishes because of (\ref{eq:bc}). Furthermore, $p > 0$ for $y \in I$, and so we conclude that $\lambda \geq 0$. This leads to the \emph{general bound}
\begin{equation}\label{eq:bound}
M^2 \geq 4\ell(4\ell+6) \ .
\end{equation}

According to the AdS/CFT dictionary, the scaling dimension of the operator dual to the bulk graviton excitation is given by the relation:\footnote{Restoring dimensions, $M^2$ is multiplied by the square of the AdS radius, which we have taken to be $1$.}
\begin{equation}\label{eq:dic}
M^2 = \Delta(\Delta-6) \ .
\end{equation}  
Thus (\ref{eq:bound}) translates into a bound on the scaling dimension
\begin{equation}\label{eq:Dbound}
	\Delta \geq 4\ell+6\ ,
\end{equation}
which yields the unitarity bound for the scalar superconformal primary of the spin-2 multiplet (see appendix \ref{multiplets}). This is a first sign that our boundary conditions are reasonable; we will analyze this further in section \ref{sub:bc}.

The spin-2 state of mass $M^2 = 4\ell(4\ell+6)$ which saturates the bound belongs to a short graviton multiplet: this the multiplet with $j=0$ in Table \ref{graviton}. The corresponding eigenfunction satisfies $S f = 0$, or
\begin{equation}
f' = \frac{c_0}{p}\ ,
\end{equation}
where $c_0$ is a constant. Since we want $f'$ to be finite at the endpoints of $I$, whereas $p$ vanishes there, $c_0$ has to be zero; in other words,
\begin{equation}
	f= \text{constant}\ .
\end{equation}
Going back to \eqref{factorY}, we conclude that
\begin{equation}\label{eq:Yshort}
	Y=\sum_{\ell=0}^\infty \sum_{m=-\ell}^\ell \beta^{\ell/2} Y_l^m
\end{equation}
for this short multiplet, which is present for all solutions in our class.

A proper Kaluza--Klein expansion, apart from the orthogonality of eigenfunctions of $S$ with different eigenvalues (proven earlier), requires that the eigenfunctions of $S$ form a complete basis for the space of functions we are considering. This is a complex task for the singular Sturm--Liouville problem at hand, and we will not attempt it. However we will see that this is the case for the examples we will analyze in section \ref{sec:ex}.

\subsection{Boundary conditions} 
\label{sub:bc}

As we have seen, it is natural to assume that $f$ and its derivative remain finite at the extrema of $I$; this led us to recover the field theory unitarity bound (\ref{eq:Dbound}). We will now check that this boundary condition on $f$ gives a proper behavior for $Y$. 

We will look at the local behavior near a regular point, and near a stack of D6-branes.\footnote{It is also possible to have O6-planes, but we will not consider them here.} These can be characterized as a single and a double zero of $\beta$ respectively. For example (\ref{R3D6}) has a single zero at $y=y_0$ and a double one at $y=-2y_0$, which corresponds to a regular point at one end and a D6-brane stack at the other.

Let us start with the behavior near a regular point, where $\beta$  has a single zero: 
\begin{equation}\label{eq:s0}
	\beta = \beta_1 (y-y_0)+ O(y-y_0)^2\ .
\end{equation}
In this case we see that there are two solutions: $f\sim$ const., and $f\sim (y-y_0)^{-\ell-\frac12}$. Since the latter diverges, according to our boundary conditions (\ref{eq:bc}) we only have to keep the solution $f\sim$ const. Going back to (\ref{factorY}), we see that $Y\sim \sum_{\ell=0}^\infty \sum_{m=-\ell}^\ell (y-y_0)^{\ell/2} Y^m_\ell$. On the other hand, plugging (\ref{eq:s0}) in (\ref{metric}) we see  \cite[Sec.~5.3]{afpt} that the local radial variable is $r=\sqrt{y-y_0}$, so locally we write $Y\sim \sum_{\ell=0}^\infty \sum_{m=-\ell}^\ell r^{\ell} Y^m_\ell$. A function of the form $r^\ell Y^m_\ell$ is indeed a smooth function, as one can check by going to local Cartesian coordinates around $r=0$.

We now turn to the case where $\beta$ has a double zero:
\begin{equation}\label{eq:d0}
	\beta = \beta_2 (y-y_0)^2+ O(y-y_0)^3\ .
\end{equation}
Now (\ref{ODE}) gives two solutions: $f\sim$ const., and $f\sim (y-y_0)^{-2\ell-1}$. Again, the latter diverges, and thus we have to keep the solution $f\sim$ const, which via (\ref{factorY}) corresponds to $Y\sim \sum_{\ell=0}^\infty \sum_{m=-\ell}^\ell (y-y_0)^{\ell} Y^m_\ell$. In this case, there is no choice of local coordinate that makes the metric regular: rather, there is a choice that turns it into a local version of a D6-brane metric \cite[Sec.~5.3]{afpt}. This is in fact simply $\rho=y-y_0$, and thus we get $Y\sim \sum_{\ell=0}^\infty \sum_{m=-\ell}^\ell \rho^{\ell} Y^m_\ell$. Again these are regular functions around $\rho=0$.

One might complain that in this case there is no reason one should expect $Y$ to be a regular function at $y=y_0$, since at this point the metric is not regular anyway. Still, for the particular massless solution (\ref{zeroF0}) (which we will analyze globally in the next section), we can use the IIA/M-theory duality; the case with a single D6-brane is known to lift to a regular point, and we can use this case as a cross-check. Locally the M-theory fibration looks like a Hopf fibration of $S^3 \subset \rr^4\to S^2 \subset \rr^3$. As is well-known, the coordinates $\{y^i : y^iy^i = 1\}$ in $\rr^3$ are quadratic in the coordinates $\{x^n : x^nx^n = 1\}$ in $\rr^4$.\footnote{The Laplacian on $S^2$ has eigenvalues $\ell(\ell+1)$, while the one on $S^3$ has eigenvalues $\tilde\ell(\tilde\ell+2)$; with $\tilde \ell=2\ell$, the two agree, once one also recalls a factor of 2 in the radius of the $S^2$ in the Hopf fibration.}  If $\rho^{\ell} Y^m_\ell$ is locally a polynomial of degree $\ell$ in the $y^i$ coordinates, it lifts to a polynomial of degree $2\ell$ in the $x^n$ coordinates, which in particular is regular.


\section{Examples}
\label{sec:ex}

We will now examine three members of the family of AdS$_7$ solutions presented in section \ref{sec:ads7}; in two cases we obtain analytic results for both the Kaluza--Klein modes and the mass spectrum. 

\subsection{The zero Romans mass solution}
\label{zeroRomansmass}

In this section we will obtain the mass spectrum of the AdS$_7$ solution with zero Romans mass. Substituting \eqref{zeroF0} for $\beta$ in \eqref{ODE} we obtain the ODE
\begin{equation}\label{Legendre}
(1-z^2) \frac{d^2u}{dz^2} - 2 z \frac{du}{dz} + \left[q(q+1) - \frac{n^2}{1-z^2}\right] u = 0 \ ,
\end{equation}
where we have introduced a new variable 
\begin{equation}
z = \frac{y}{y_0}\ , \qquad z \in [-1,1] \ ,
\end{equation}
and function
\begin{equation}
u = (1-z^2)^{\ell+1/2} f \ .
\end{equation}
Furthermore, the parameters $q$ and $n$ are related to $M^2$ and $\ell$ as
\begin{equation}\label{parmtrs}
q(q+1) = \frac{M^2}{4} + 2 \ , \qquad n = 2\ell + 1 \ .
\end{equation}
Equation \eqref{Legendre}  is the associated Legendre differential equation with general solution
\begin{equation}
u(z) = C_1 P_q^n(z) + C_2  Q_q^n(z) 
\end{equation} 
where $P_q^n$ and $Q_q^n$ are the associated Legendre functions of first and second kind respectively; $C_1$, $C_2$ are arbitrary constants.

We want to impose that $f \propto (1-z^2)^{-\ell-1/2} u$ in \eqref{factorY} is regular at $z = \pm 1$. In order for this to be the case, $u$ needs to be regular; this is true for $C_2 = 0$ and 
\begin{equation}
\{q,n \in \mathbb{Z} \, : \, q \geq n \geq 0 \} \ ,
\end{equation}
in which case $P_q^{n}$ are the associated Legendre polynomials. Since $\ell \geq 0$ and hence $n \geq 1$, in fact $q \geq 1$. Finally,  
\begin{equation}
P_q^n(z) = (-1)^n (1-z^2)^{n/2} \frac{d^n}{dz^n} P_q(z) = -  (1-z^2)^{\ell+1/2} \frac{d^n}{dz^n} P_q(z)
\end{equation}
where $P_q$ are the Legendre polynomials, and so 
\begin{equation}
f \propto  (1-z^2)^{-\ell-1/2} P_q^n = -\frac{d^n}{dz^n} P_q(z) \ ,
\end{equation}
is regular at $z=\pm1$.

To obtain the mass spectrum, we invert \eqref{parmtrs}:
\begin{equation}\label{eq:spec-massless}
M^2 = 2\tilde \jmath(2\tilde \jmath+6) \ ,  \qquad 
\{\tilde \jmath \equiv q - 1 \in \mathbb{Z} \, : \, \tilde \jmath \geq 2 \ell \geq 0 \} \ .
\end{equation}
As expected, this the same spin-2 mass spectrum as that of eleven-dimensional supergravity on $S^4$ \cite{vanNieuwenhuizen:1984iz}.\footnote{\label{foot:fac4}A factor of $4$ difference is due to a different radius for AdS$_7$, which in \cite{vanNieuwenhuizen:1984iz} is taken to be $2$ ($M^2$ is multiplied by the square of the AdS$_7$ radius when dimensions are restored).}
The inequality $q \geq n$ or $\tilde \jmath \geq 2\ell$ becomes (\ref{eq:bound}).

Taking the associated Legendre polynomials as a basis for functions in the interval $I$, we can write the full Kaluza--Klein expansion of the metric perturbation $\delta \bar{g}_{\mu\nu}$ as 
\begin{equation}
\delta \bar{g}_{\mu\nu} = \sum_{\ell=0}^{\infty} \sum_{\tilde\jmath=2\ell}^{\infty} \sum_{m=-\ell}^\ell h^{\tilde\jmath,\ell,m}_{\mu\nu} \beta^{\ell/2}  P^{(2\ell+1)}_{(\tilde\jmath+1)} Y_\ell^m \ .
\end{equation}

\subsection{The $\mathbb{R}^3$--D6 solution}
\label{sub:R3D6}
In this section we determine the mass spectrum for the $\mathbb{R}^3$--D6 solution characterized by \eqref{R3D6}. Recall that  $y_0 < 0$ and $y_0 \leq y \leq -2 y_0$. After a change of variables
\begin{equation}
z = \frac{1}{3} \left( 2 + \frac{y}{y_0} \right) \ , \qquad z \in [0,1] \ ,
\end{equation}
\eqref{ODE} becomes the hypergeometric differential equation \eqref{hyper}; in the main text we will denote its solution (\ref{eq:hyperF}) by ${}_2F_1\equiv F$. The parameters $a$, $b$ and $c$ are determined by $\ell$ and $M^2$ via
\begin{equation}\label{eq:abc}
a \equiv \frac{1}{4} (6\ell + 5 - \tau) \ , \qquad
b \equiv  \frac{1}{4} (6\ell + 5 + \tau) \ , \qquad
c \equiv 2(\ell + 1) \ , 
\end{equation}
where
\begin{equation}
\tau^2 \equiv (6\ell + 5)^2 + 3\left[M^2 - 4\ell(4\ell+6)\right]\ .
\end{equation}
 In the neighborhood of $z=0$ the general solution would be a linear combination of 
$f_1 = F(a,b;c;z)$ and $f_2 = z^{1-c}F(a-c+1,b-c+1;2-c;z)$ if $2-c=-2\ell$ was not a non-positive integer. Since for us $2 - c \leq 0$, $f_2$ is replaced by a solution with more complicated expression; see for example \cite[Sec.~15.10]{NIST}.

Next we need to impose that $f$ is regular. 
At $z=0$ only $f_1$ is regular and hence
\begin{equation}\label{eq:fF}
f = C \, F(a,b;c;z) \ , \qquad C={\rm const.}
\end{equation}
To check regularity at $z=1$ we employ \eqref{glueeq} which is valid for the case at hand since $a+b-c = \ell+\frac{1}{2}$ is not an integer. The relation \eqref{glueeq} shows that $f$ has a pole of order $c-a-b = -\ell-\frac{1}{2}$. In order to eliminate the pole we need to take $a$ or $b$ to be a non-positive integer, at which case $f$ truncates to a polynomial. 

Imposing that $a =-j$, $j \in \mathbb{Z}_{\geq 0}$, fixes the mass $M^2$ in terms of $j$ and $\ell$. We find the following mass spectrum:
\begin{equation}\label{eq:spectrum-R3D6}
\begin{split}
	M^2 &= (2j+4\ell)(2j+4\ell+6) + \frac{1}{3}2j(2j+2) \\
		&= 4\ell(4\ell+6) + \frac83j(5+2j+6\ell)\ ,
\end{split}
\end{equation}
in agreement with the bound (\ref{eq:bound}).

Looking back at $f$
\begin{equation}
f = C \, F(-j,j+3\ell+\tfrac{5}{2};2l+2;z) \ ,
\end{equation}
we see that it has become a Jacobi polynomial \eqref{JPoly} for $x= 1-2z$:
\begin{equation}
 f = C \, \frac{(2\ell+1)!\,n!}{(2\ell+1+n)!} P^{(2\ell+1,\ell+\tfrac{1}{2})}_j(1-2z) \ .
\end{equation}
Note that the weighted inner product introduced in \eqref{innerp} with weight function $w = \beta^{\ell+1/2} \propto  z^{2\ell+1} (1-z)^{\ell+\tfrac{1}{2}}$ gives the weighted inner product with respect to which the Jacobi polynomials are orthogonal, as expected from the general analysis of section \ref{sub:bound}.

Taking the Jacobi polynomials as a basis for functions in the interval $I$, we can write the full Kaluza--Klein expansion of the metric perturbation $\delta \bar{g}_{\mu\nu}$ as 
\begin{equation}
\delta \bar{g}_{\mu\nu} = \sum_{j=0}^{\infty} \sum_{\ell=0}^{\infty} \sum_{m=-\ell}^\ell h^{j,\ell,m}_{\mu\nu} \beta^{\ell/2}  P^{(2\ell+1,\ell+\tfrac{1}{2})}_j Y_\ell^m \ .
\end{equation}

\subsection{Solutions with D8-branes} 
\label{sub:d8}

We will now look at solutions with D8-branes. We will illustrate the procedure by focusing on the solution with a single stack of D8-branes reviewed briefly in section \ref{sec:ads7}. 

Since this solution is obtained by gluing two pieces of the $\rr^3$--D6 solution, we can borrow some of the analysis in section \ref{sub:R3D6}. However, there are some changes in the regularity analysis, due to the different global structure of the solution and to the presence of the D8-brane stack. 

We will cover the solution with two coordinate patches. One has coordinate  $z = \frac13\left(2+\frac y{y_0}\right)$ with $z\in [z_{\rm D8},1]$, where $z_{\rm D8}=\frac13\left(2+\frac {y_{\rm D8}}{y_0}\right)$. The other patch has coordinate $z' = \frac13\left(2+\frac y{y_0'}\right)$, which similarly is defined in the interval $z'\in [z'_{\rm D8},1]$, where $z'_{\rm D8}=\frac13\left(2+\frac {y_{\rm D8}}{y_0'}\right)$. Using (\ref{eq:1D8}) we can compute
\begin{equation}
	z_{\rm D8}=\frac{2\mu}{N+\mu} \ ,\qquad z'_{\rm D8} = \frac{2(N-\mu)}{2N-\mu}\ .
\end{equation}
The two coordinates are related by the transition function 
\begin{equation}
	z'=\frac13\left(2 +\frac{y_0}{y'_0}(3z-2)\right)=\frac{z(N+\mu)-2N}{\mu-2N}\ .
\end{equation}
In these coordinates, the north and south pole of the geometry (which are both regular) correspond to $z=1$ and $z'=1$. 

In both charts, \eqref{ODE} becomes again the hypergeometric differential equation \eqref{hyper}. However, in section \ref{sub:R3D6} we started by imposing regularity around $z=0$ (which led to (\ref{eq:fF})), while here $z=0$ is not included in either chart. On the other hand, $z=1$ is present in both charts; thus we can start our analysis by imposing regularity around that point. The general solution to (\ref{hyper}) around $z=1$ is a linear combination of $F(a,b;a+b-c+1;1-z)$ and $(1-z)^{c-a-b}F(c-a,c-b;c-a-b+1;1-z)$, with the constants given in (\ref{eq:abc}); see \cite[Sec.~15.10]{NIST}. Since $c-a-b=-\ell-\frac12$, the second solution diverges near $z=1$, and must be discarded. Thus in the first chart we have the solution $C \, F(z) \equiv  C\, F(a,b;\ell+ 3/2;1-z)$, where $C$ is an arbitrary constant, and on the second one the solution  $C' \, F(z') \equiv  C'\, F(a,b;\ell+ 3/2;1-z')$, with a different constant $C'$.

We now have to impose an appropriate condition at the locus where the D8-brane stack is present. This can be argued as follows. If either $f$ or $df/dy$ were discontinuous at the D8-brane locus, a $\delta$ function or its derivative would appear on the right-hand side of (\ref{ODE}). However, as we remarked already in section \ref{sub:mass}, (\ref{ODE}) is not altered by any contribution to the stress-energy tensor $T_{\mu \nu}$, not even by localized sources. Thus we conclude that $f$ and $df/dy$ should be continuous at the D8-brane locus. This is formally similar to how one solves a Schr\"odinger problem with a discontinuous potential.

The equations we need to impose now read $C \, F(z)|_{z=z_{\rm D8}}=C' \, F(z')|_{z'=z'_{\rm D8}}$ and $C \, (d F(z)/dy)|_{z=z_{\rm D8}}=C' \, (dF(z')/dy)|_{z'=z'_{\rm D8}}$. The system admits a solution if and only if the matrix
$\left(\begin{smallmatrix}
F(z) & F(z') \\ \\ dF(z)/dy \ & \ dF(z')/dy
\end{smallmatrix}
\right)\Big|_{y=y_{\rm D8}}
$
has zero determinant.  This leads to the condition
\begin{equation}
	\frac{d z}{d z'} \frac{d}{dz} \log F(a,b;\ell+3/2;1-z)|_{z=z_{\rm D8}}=
	\frac{d}{dz'} \log F(a,b;\ell+3/2;1-z')|_{z'=z'_{\rm D8}},
\end{equation}
which can be solved numerically by varying $\tau$ in (\ref{eq:abc}). 

Empirically we find that the allowed values for $M^2$ in $\tau$ are well approximated by
\begin{equation}
	M^2=4\ell(4\ell+6)+ j (\alpha_0 + \alpha_1 j + \alpha_2 \ell) ,
\end{equation}
where the $\alpha_i$'s depend on $\mu/ N$ only, whereas $j\in \zz_{\ge 0}$. This is qualitatively similar to what we obtained in (\ref{eq:spectrum-R3D6}) for the $\rr^3$--D6 solution.


\section{Field theory interpretation}
\label{sec:ft}

The $(1,0)$ SCFT's dual to the AdS$_7$ type IIA supergravity solutions do not have a Lagrangian description. Nevertheless, in this section we will translate the results obtained so far in a field-theoretic language, and at least attempt an interpretation of these in terms of the effective field theory descriptions of the CFT's.

\begin{figure}[ht]
\centering	
	\subfigure[\label{fig:massless}]{\includegraphics[width=6cm]{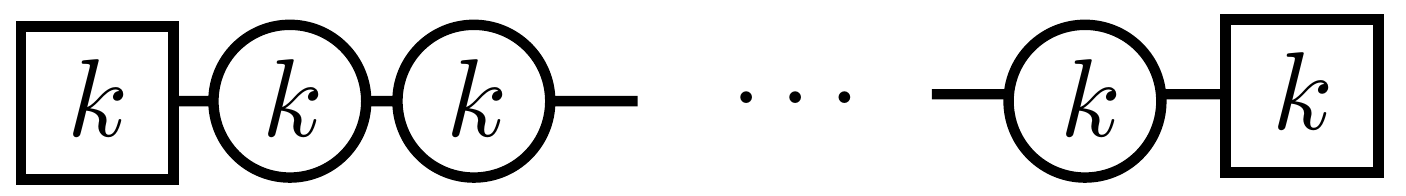}}\hspace{2cm}
	\subfigure[\label{fig:R3-D6}]{\includegraphics[width=6cm]{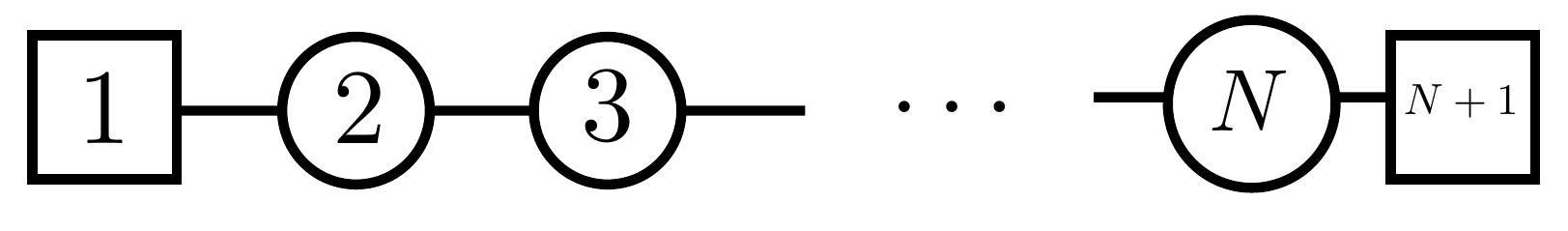}}\\
	\subfigure[\label{fig:1D8}]{\includegraphics[width=7cm]{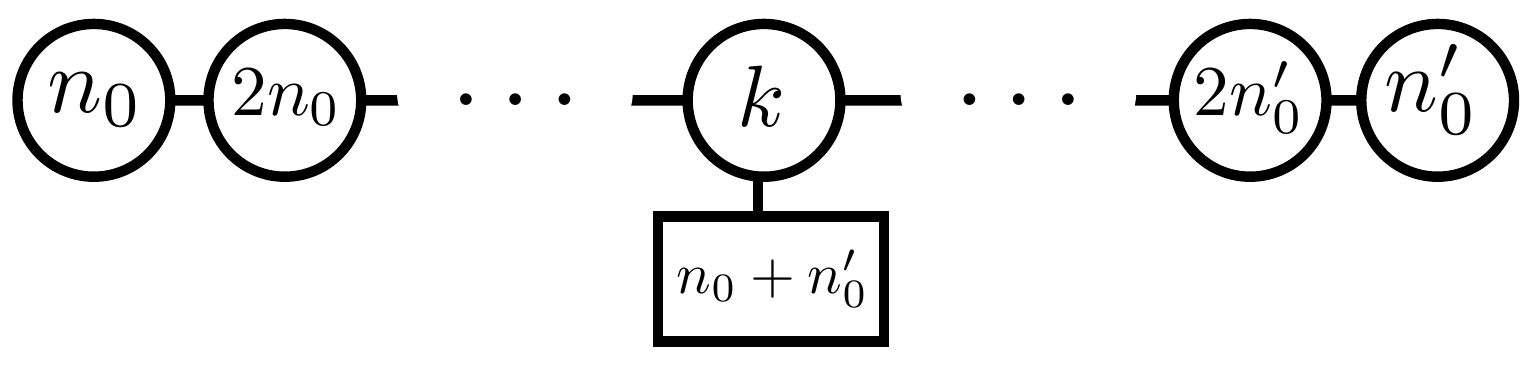}}\\
	\caption{\small Some examples of linear quivers. Round nodes denote gauge symmetries and square nodes denote flavor symmetries. Links correspond to (bi)fundamental hypermultiplets, with tensor multiplets on horizontal ones.  \subref{fig:massless} corresponds to the massless solution of section \ref{zeroRomansmass}; \subref{fig:R3-D6} to the $\rr^3$--D6 solution of section \ref{sub:R3D6}; \subref{fig:1D8} to the solution with one D8-brane stack in section \ref{sub:d8}. In particular, for the latter one, $k=n_0(N- \mu)=n_0' \mu$, where $n_0=2\pi F_0$, $n_0'=-2\pi F_0'$.}
	\label{fig:cft}
\end{figure}

These effective theories consist of a chain of SU$(r_i)$ vector multiplets, $i=1,\ldots,N-1$. Simplifying a bit, these are coupled to hypermultiplets in the bifundamental representation $\overline{\mathbf{r_i}}\otimes \mathbf{r_{i+1}}$,  $f_i=2r_i-r_{i+1}-r_{i-1}$ hypermultiplets in the fundamental representation $\mathbf{r_i}$, and $N$ tensor multiplets. (See Figure \ref{fig:cft} for some examples.) Let us call $h_i^I$ the scalars in the bifundamental hypermultiplets ($I$ being an SU(2)$_{\rm R}$ index) and $\Phi_i$ the scalars in the tensor multiplets. The gauge kinetic terms are of the form $(\Phi_{i+1} - \Phi_{i}){\rm Tr}|F_i|^2$; when all the $\Phi_i$ coincide, the gauge couplings are divergent, and the theory is strongly coupled. This is the point which is supposed to correspond to a CFT.  

A result of this paper is that all the CFT's under study, irrespectively of the choice of the $r_i$, have a short spin-2 multiplet with dimension $\Delta=4\ell+6$, where $\ell$ is the SU(2)$_{\rm R}$ spin; this saturates the bound (\ref{eq:bound}). The structure of the multiplet is given in Table \ref{graviton} for $j=0$. The scalar superconformal primary in the multiplet has dimension 
\begin{equation}\label{eq:univ}
	\Delta_{\rm pr}= 4 \ell + 4 \ .
\end{equation}

There are two theories for which we have found the full spectrum of spin-2 operators, short and long; these are dual to the solutions discussed in sections \ref{zeroRomansmass} and \ref{sub:R3D6}. The first can be lifted to M-theory, and there it can be thought of as a $\zz_k$ orbifold of the theory living on a stack of M5-branes; its effective description is a quiver depicted in Figure \ref{fig:massless}. From (\ref{eq:spec-massless}) and (\ref{eq:dic}) it follows that the spin-2 operators have dimension
\begin{equation}
	\Delta= 4 \ell + 2j + 6 \ ,
\end{equation}
where $j\equiv \tilde \jmath -2 \ell \in \mathbb{Z}_{\geq 0}$ is introduced.
For $j=0$ and $j=1$, the operators belong to short multiplets; see the discussion in appendix \ref{multiplets}.
The second theory is the one whose effective description is depicted in Figure \ref{fig:R3-D6}. From  (\ref{eq:spectrum-R3D6}) and (\ref{eq:dic})  it follows that the spin-2 operators have dimension 
\begin{equation}\label{eq:sc-sp}
	\Delta = 3 + \sqrt{(4 \ell+2j+3)^2 + \frac13 2 j (2j+2)} \ .
\end{equation}
For $j=0$, the operator belongs to a short multiplet.

More general theories involve D8-branes; in the effective description, these correspond to the presence of non-zero $f_i=2 r_i - r_{i-1}-r_{i+1}$. Unfortunately in such cases it is hard to solve for the spectrum explicitly, although in section \ref{sub:d8} we have described how one can obtain such information numerically. Preliminary evidence seems to point to modification of the coefficients in (\ref{eq:sc-sp}). 

Let us now go back to the universal short multiplets in (\ref{eq:univ}). As anticipated in the introduction, it is tempting to match them in the effective field theory description with a bound state of $2\ell$ hypermultiplets and of the stress-energy tensor multiplet ${\cal T}$. There are several gauge invariant operators that one can obtain by tracing hypermultiplets; for example one can write
\begin{equation}\label{eq:tr}
	{\rm Tr}\left(h^{(I_1}_i h^{I_2\,\dagger}_i\ldots h^{I_{2\ell-1}}_i h^{I_{2\ell})\,\dagger}_i\right) T_{\alpha \beta} \ ,
\end{equation}
where $T_{\alpha\beta}$ is the stress-energy tensor.

The versions for different $i$ of this operator are all related to each other by the D-term and F-term equations, but the relations in general also involve the fundamentals $f_i$. More general operators can also be related to (\ref{eq:tr}) via D-term and F-term equations, again in general involving the $f_i$. The $f_i$ are related to the presence of D-brane stacks in the gravity solution; thus it is natural to conjecture that the operators corresponding to the massive spin-2 excitations we analyzed in this paper should not involve them, and that they should rather be a combination of the (\ref{eq:tr}). 

The operators (\ref{eq:tr}) are spin-2 operators, and have classical dimension $4\ell+6$. In a superconformal theory, $T_{\alpha \beta}$ belongs to a multiplet which has a scalar superconformal primary $\phi$, with dimension $4$; correspondingly, the scalar operator in the short spin-2 multiplets we have considered will be obtained by replacing $T_{\alpha \beta}\to \phi$ in (\ref{eq:tr}). This has classical dimension equal to $4\ell+4$, matching (\ref{eq:univ}). For the SCFT of a single tensor multiplet, $\phi$ is $\frac{1}{2}\Phi^2$ \cite{howe-sierra-townsend}. Operators belonging to long multiplets could be described by a product of \eqref{eq:tr} with $\Phi^j$.

\section*{Acknowledgements}
We would like to thank F.~Benini, C.~Cordova, D.~Rodr\'{\i}guez-G\'{o}mez and A.~Zaffaroni for interesting discussions. We are supported in part by INFN and by the European Research Council under the European Union's Seventh Framework Program (FP/2007-2013) -- ERC Grant Agreement n. 307286 (XD-STRING). A.T.~is also supported by the MIUR-FIRB grant RBFR10QS5J ``String Theory and Fundamental Interactions''.

\appendix

\section{\OSp($6,2|1$) multiplets}
\label{multiplets}

In this appendix we present the \OSp($6,2|1$) supermultiplets corresponding to supergravity fields. One way to construct these is to use the oscillator method analogously to \cite{Gunaydin:1984wc}, where it was employed for \OSp($6,2|2$) and the spectrum of eleven-dimensional supergravity on $S^4$. 
Such a construction was carried out for the vector multiplet and a short graviton multiplet of \OSp($6,2|1$) in \cite{Gimon:1999yu}. 

We have taken a different approach, and obtained the \OSp$(6,2|1)$ supermultiplets by decomposing the \OSp$(6,2|2)$ supermultiplet related to the reduction of eleven-dimensional supergravity on $S^4$, summarized for example in \cite[Table 4]{Beccaria:2015ypa}. (See also footnote \ref{foot:clay}.) We branch the representations of the $\mathfrak{sp}(2)$ R-symmetry algebra of \OSp($6,2|2$) into representations of an \SU$(2)_1$ $\oplus$ \SU$(2)_2$ subalgebra,\footnote{We have used that the $\mathfrak{sp}(2) \simeq$ \SO$(5)$ representation with Dynkin labels $[q, p]$ branches as $\sum_{n_1=0}^q \sum_{n_2=0}^p ({\bf q+p+1-n_1-n_2}, {\bf p+1+n_1-n_2})$.} and take \SU$(2)_2$ to be the R-symmetry algebra of \OSp$(6,2|1)$. Members of the supermultiplet that transform in the same representation of \SU$(2)_1$ form \OSp($6,2|1$) supermultiplets. In particular the  {\bf p+1} representation of  \SU$(2)_1$ yields the vector multiplet, the {\bf p} representation the gravitino multiplet and the {\bf p$-$1$-$j} representation, where $j$ is an integer greater than or equal to $0$, the graviton multiplet. (A bold font is used to denote the dimensions of the \SU$(2)$ representations.) 

In Tables \ref{vector}--\ref{graviton}, the numbers $(\Delta;\,\omega_1,\,\omega_2,\,\omega_3)$ characterize the \SO$(6,2)$ representations: $\Delta$ is the scaling dimension and $(\omega_1,\,\omega_2,\,\omega_3)$ the \SO$(6)$ highest weights.\footnote{The are related to Dynkin labels as follows: $D_1 = \omega_2-\omega_3$, $D_2 = \omega_1-\omega_2$, $D_3 = \omega_2+\omega_3$.} ${\bf p}$ denotes the dimension of an \SU$(2)_{\rm R}$ representation and $\ell = (p-1)/2$ the corresponding spin. The superconformal primaries of the multiplets are scalars, and according to \cite{Minwalla:1997ka,dobrev} unitary representations whose scalar superconformal primary has scaling dimension $\Delta_{\rm pr} = 4\ell$, $4\ell+2$, $4\ell+4$ or $4\ell+6$ are short. For the rest of the unitary representations $\Delta_{\rm pr} \geq 4\ell+6$. In this sense the vector, gravitino, and graviton multiplets with $j=0$ or $1$ are short.

Let us mention that a superfield formalism is also available \cite{Ferrara:2000xg}. This can be used for constructing unitary representations of \OSp$(6,2|1)$, and it provides an organizing principle for operators. The construction of the representations proceeds by tensoring supersingletons described by constrained superfields; for the $(1,0)$ superconformal algebra these are: (i) $W$, whose components form an $(1,0)$ tensor multiplet and (ii) $W^I$, where $I$ is an \SU(2)$_{\rm R}$ index, whose components form a hypermultiplet. Tensor products of these can be used to construct the vector, gravitino and graviton multiplets. For example the massless graviton multiplet or stress-energy tensor multiplet can be constructred out of the product of two tensor multiplets $(W)^2$. This formalism is related to the ``place-holder fields'' fomalism used in \cite{Gimon:1999yu}.

\begin{table}[ht]
\centering
\begin{tabular}{|c|l|c|}
\hline
$p\geq$ & $(\Delta;\omega_1,\omega_2,\omega_3)$ & \SU$(2)_{\rm R}$ \\ \hline \hline
\multirow{2}{*}{$1$} & $(2p;0,0,0)$ & {\bf p$+$1} \\ \cline{2-3}
           & $(2p+\frac{1}{2};\frac{1}{2},\frac{1}{2},\frac{1}{2})$ & {\bf p} \\ 
\hline
$2$ & $(2p+1;1,0,0)$ & {\bf p$-$1} \\ \hline 
$3$ & $(2p+\frac{3}{2};\frac{1}{2},\frac{1}{2},-\frac{1}{2})$ & {\bf p$-$2} \\ \hline 
$4$ & $(2p+2;0,0,0)$ & {\bf p$-$3} \\ \hline
\end{tabular}
\caption{Vector Multiplet ($p\geq2$). The superconformal primary has $\Delta_{\rm pr} = 4 \ell$. $p=1$ yields the hypermultiplet. $p=2$ yields the massless vector multiplet.}
\label{vector}
\end{table}

\begin{table}[ht]
\centering
\begin{tabular}{|c|l|c|}
\hline
$p\geq$ & $(\Delta;\omega_1,\omega_2,\omega_3)$ & \SU$(2)_{\rm R}$ \\ \hline\hline
\multirow{3}{*}{$1$} & $(2p;0,0,0)$ & {\bf p} \\ \cline{2-3}
& $(2p+\frac{1}{2};\frac{1}{2},\frac{1}{2},\frac{1}{2})$ & {\bf p$+$1} \\ \cline{2-3} 
& $(2p+1;1,1,1)$ & {\bf p} \\ \hline
\multirow{3}{*}{$2$}& $(2p+\frac{1}{2};\frac{1}{2},\frac{1}{2},\frac{1}{2})$ & {\bf p$-$1} \\ \cline{2-3}
& $(2p+1;1,0,0)$ & {\bf p} \\ \cline{2-3}
& $(2p+\frac{3}{2};\frac{3}{2},\frac{1}{2},\frac{1}{2})$& {\bf p$-$1} \\ \hline
\multirow{3}{*}{$3$} & $(2p+1;1,0,0)$ & {\bf p$-$2} \\ \cline{2-3}
& $(2p+\frac{3}{2};\frac{1}{2},\frac{1}{2},-\frac{1}{2})$ & {\bf p$-$1} \\ \cline{2-3}
& $(2p+2;1,1,0)$ & {\bf p$-$2} \\ \hline 
\multirow{3}{*}{$4$} &  $(2p+\frac{3}{2};\frac{1}{2},\frac{1}{2},-\frac{1}{2})$ & {\bf p$-$3} \\ \cline{2-3}
& $(2p+2;0,0,0)$ & {\bf p$-$2} \\ \cline{2-3}
& ($2p+\frac{5}{2};\frac{1}{2},\frac{1}{2},\frac{1}{2})$ & {\bf p$-$3} \\ \hline
$5$ & $(2p+2;0,0,0)$ & {\bf p$-$4} \\ \hline
\end{tabular}
\caption{Gravitino Multiplet ($p\geq2$). The superconformal primary has $\Delta_{\rm pr} = 4\ell+2$. $p=1$ yields the tensor multplet. $p=2$ yields the massless gravitino multiplet.}
\label{gravitino}
\end{table}

\begin{table}[t]
\begin{minipage}[t]{.5\linewidth}\centering
\begin{tabular}{|c|c|l|c|}
\hline
$p\geq$ & $j\geq$ & $(\Delta;\omega_1,\omega_2,\omega_3)$ & \SU$(2)_{\rm R}$ \\ \hline\hline
\multirow{15}{*}{$j+2$} 
&\multirow{6}{*}{$0$}
& $(2p;0,0,0)$ & {\bf p$-$j$-$1} \\ \cline{3-4}
&& $(2p+\frac{1}{2}; \frac{1}{2},\frac{1}{2},\frac{1}{2})$ & {\bf p$-$j} \\ \cline{3-4} 
& & $(2p+1;1,1,1)$ & {\bf p$-$j$-$1}\\ \cline{3-4} 
 && $(2p+1;1,0,0)$ & {\bf p$-$j$+$1} \\ \cline{3-4}
& & $(2p+\frac{3}{2};\frac{3}{2},\frac{1}{2},\frac{1}{2})$ & {\bf p$-$j} \\ \cline{3-4}
& & $(2p+2;2,0,0)$ & {\bf p$-$j$-$1} \\ \cline{2-4} 
& \multirow{4}{*}{$1$} & $(2p+\frac{3}{2};\frac{1}{2},\frac{1}{2},-\frac{1}{2})$ & {\bf p$-$j$+$2} \\ \cline{3-4}
& & $(2p+2;1,1,0)$ & {\bf p$-$j$+$1} \\ \cline{3-4}
& & $(2p+\frac{5}{2};\frac{3}{2},\frac{1}{2},-\frac{1}{2})$ & {\bf p$-$j} \\ \cline{3-4}
& & $(2p+3;1,1,-1)$ & {\bf p$-$j$-$1} \\ \cline{2-4}
& \multirow{5}{*}{$2$} & $(2p+2;0,0,0)$ & {\bf p$-$j$+$3} \\ \cline{3-4}
 && $(2p+\frac{5}{2};\frac{1}{2},\frac{1}{2},\frac{1}{2})$ & {\bf p$-$j$+$2} \\ \cline{3-4}
& & $(2p+3;1,0,0)$ & {\bf p$-$j$+$1} \\ \cline{3-4}
& & $(2p+\frac{7}{2};\frac{1}{2},\frac{1}{2},-\frac{1}{2})$ &
 {\bf p$-$j} \\ \cline{3-4} 
& & $(2p+4;0,0,0)$ & {\bf p$-$j$-$1} 
\\ \hline 
\multicolumn{1}{c}{} \\
\multicolumn{1}{c}{} \\
\multicolumn{1}{c}{} \\
\multicolumn{1}{c}{} \\
\multicolumn{1}{c}{} \\
\end{tabular}
\end{minipage}
\hspace{.5cm}
\begin{minipage}[t]{0.5\linewidth}\centering
\begin{tabular}{|c|c|l|c|}
\hline
$p\geq$& $j\geq$& $(\Delta;\omega_1,\omega_2,\omega_3)$ & \SU$(2)_{\rm R}$ \\ \hline\hline
\multirow{10}{*}{\rotatebox{0}{$j+3$}} & \multirow{6}{*}{\rotatebox{0}{$0$}} & $(2p+\frac{1}{2};\frac{1}{2},\frac{1}{2},\frac{1}{2})$ & {\bf p$-$j$-$2} \\ \cline{3-4}
& & $(2p+1; 1,0,0)$ & {\bf p$-$j$-$1} \\ \cline{3-4} 
& & $(2p+\frac{3}{2};\frac{3}{2},\frac{1}{2},\frac{1}{2})$ & {\bf p$-$j$-$2}\\ \cline{3-4} 
& & $(2p+\frac{3}{2};\frac{1}{2},\frac{1}{2},-\frac{1}{2})$ & {\bf p$-$j} \\ \cline{3-4}
& & $(2p+2;1,1,0)$ & {\bf p$-$j$-$1} \\ \cline{3-4}
& & $(2p+\frac{5}{2};\frac{3}{2},\frac{1}{2},-\frac{1}{2})$ & {\bf p$-$j$-$2} \\ \cline{2-4}
& \multirow{4}{*}{\rotatebox{0}{$1$}} & $(2p+2;0,0,0)$ & {\bf p$-$j$+$1} \\ \cline{3-4}
& & $(2p+\frac{5}{2};\frac{1}{2},\frac{1}{2},\frac{1}{2})$ & {\bf p$-$j} \\ \cline{3-4}
& & $(2p+3;1,0,0)$ & {\bf p$-$j$-$1} \\ \cline{3-4}  
& & $(2p+\frac{7}{2};\frac{1}{2},\frac{1}{2},-\frac{1}{2})$ & {\bf p$-$j$-$2} \\ \hline 
\multirow{6}{*}{\rotatebox{0}{$j+4$}} &\multirow{6}{*}{\rotatebox{0}{$0$}} & $(2p+1;1,0,0)$ & {\bf p$-$j$-$3} \\ \cline{3-4}
& & $(2p+\frac{3}{2}; \frac{1}{2},\frac{1}{2},-\frac{1}{2})$ & {\bf p$-$j$-$2} \\ \cline{3-4} 
& & $(2p+2;1,1,0)$ & {\bf p$-$j$-$3}\\ \cline{3-4} 
& & $(2p+2;0,0,0)$ & {\bf p$-$j$-$1}\\ \cline{3-4}
& & $(2p+\frac{5}{2};\frac{1}{2},\frac{1}{2},\frac{1}{2})$ & {\bf p$-$j$-$2} \\ \cline{3-4}
& & $(2p+3;1,0,0)$ & {\bf p$-$j$-$3} \\ \hline  
\multirow{3}{*}{\rotatebox{0}{$j+5$}} & \multirow{3}{*}{\rotatebox{0}{$0$}} &$(2p+\frac{3}{2};\frac{1}{2},\frac{1}{2},-\frac{1}{2})$ & {\bf p$-$j$-$4} \\ \cline{3-4}
& & $(2p+2; 0,0,0)$ & {\bf p$-$j$-$3} \\ \cline{3-4} 
& & $(2p+\frac{5}{2};\frac{1}{2},\frac{1}{2},\frac{1}{2})$ & {\bf p$-$j$-$4} \\ \hline 
$j+6$ & $0$ &  $(2p+2;0,0,0)$ & {\bf p$-$j$-$5}  \\ \hline
\end{tabular}
\end{minipage}
\vspace{.5cm}
\caption{Graviton Multiplet. The superconformal primary has $\Delta_{\rm pr} = 4\ell + 2 j+4$. For $j=0$ or $1$ the multiplet is short. $j=0$, $p=2$ yields the massless graviton multiplet which forms the minimal gauged supergravity in seven dimensions \cite{Townsend:1983kk}.} 
\label{graviton}
\end{table}

\section{\texorpdfstring{Non-supersymmetric AdS$_7$ solutions}{Non-supersymmetric AdS(7) solutions}}
\label{non-susy}

Given the truncation of \cite{prt}, the non-supersymmetric anti-de Sitter vacuum of the minimal gauged supergravity in seven dimensions can be uplifted to a family of non-supersymmetric AdS$_7$ solutions of massive type IIA supergravity, which are in one-to-one correspondence with the supersymmetric ones. Their metric, in string frame, reads
\begin{equation}\label{nonsusy-metric}
ds^2_{10} = \frac{e^{2A} }{\sqrt{2}}\left( \frac{3}{2}ds^2_{\mathrm{AdS}_7} - \frac{1}{16} \frac{\beta'}{y\beta} \, dy^2 + \frac{1}{4}\frac{\beta}{2 \beta-y \beta'} ds^2_{S^2} \right) \ ,
\qquad
e^{2A}= \frac49 \left(-\frac{\beta'}{y}\right)^{1/2} \ .
\end{equation}
The AdS$_7$ metric is of radius one. The dilaton is given by
\begin{equation}
e^{2\phi} = \frac{1}{144\sqrt{2}}\frac{(-\beta'/y)^{5/2}}{2 \beta - y \beta'} \ .
\end{equation}
These solutions were recently interpreted holographically in \cite{apruzzi-dibitetto-tizzano}.

The mass spectrum of the spin-2 excitations is determined by the eigenvalues of the differential operator $\LL$ defined in \eqref{Lope}. Its expression for the non-supersymmetric solutions can be linked to the one for the supersymmetric solutions as
\begin{equation}
\LL_{\rm non-susy} = \frac{3}{2} \LL_{\rm susy} - 12 \Delta_{S^2} \ ,
\end{equation}
where $\Delta_{S^2}$ is the $S^2$ Laplacian. Consequently, --- taking into account the expansion \eqref{factorY} --- 
the mass spectra for the non-supersymmetric and supersymmetric AdS$_7$ solutions are related via  
\begin{equation}
M^2_{\rm non-susy} = \frac{3}{2} M^2_{\rm susy} - 12 \ell(\ell+1) \ .
\end{equation}
Given that $M^2_{\rm susy} \geq 4\ell(4\ell+6)$, it follows that $M^2_{\rm non-susy}$ is positive, and hence the unitarity bound for a spin-2 field is satisfied (see for example \cite[(3.46),(3.47)]{rahman-taronna}).

\section{Hypergeometric function and Jacobi polynomials}
\label{hyperJac}

\subsubsection*{Hypergeometric function}

The hypergeometric function is defined for $|z| < 1$ by the power series
\begin{equation}\label{eq:hyperF}
{}_2F_1(a,b;c;z) = \sum_{k=0}^\infty \frac{(a)_k(b)_k}{(c)_k} \frac{z^k}{k!}\ ,
\end{equation}
where $z$ is a complex variable, and $a$, $b$, $c$, are parameters which can take
arbitrary real or complex values, provided that $c \neq 0, - 1, - 2,  \dots$.
The symbol $(n)_k$ denotes 
\begin{equation}
(n)_k = n(n+1) \dots (n+k-1) \ , \qquad k>0\ , 
\end{equation}
 with $(n)_0 = 1$.
The hypergeometric function becomes a polynomial whenever $a$ or $b$ is a non-positive integer.

For $\R(c-a-b) > 0$,
\begin{equation}
\lim_{z\to 1^-} {}_2F_1(a,b;c;z) = \frac{\Gamma(c)\Gamma(c-a-b)}{\Gamma(c-a)\Gamma(c-b)}\ .
\end{equation}

The hypergeometric function is a solution of the hypergeometric differential equation
\begin{equation}\label{hyper}
z(1-z) \frac{d^2u}{dz^2} +[c-(a+b+1)z]\frac{du}{dz} - a b u = 0\ ,
\end{equation}
which has regular singular points at $z=0,1,\infty$. In the neighborhood of $z=0$ the general solution is a linear combination of 
${}_2F_1(a,b;c;z)$ and $z^{1-c}{}_2F_1(a-c+1,b-c+1;2-c;z)$.   In the neighborhood of $z=1$ the general solution is a linear combination of ${}_2F_1(a,b;a+b-c+1;1-z)$ and $(1-z)^{c-a-b}{}_2F_1(c-a,c-b;c-a-b+1;1-z)$.  

Provided that $a+b-c$ is not an integer, the following relation holds
\begin{equation}\label{glueeq}
\begin{split}
{}_2F_1(a,b;c;z) = &\ \frac{\Gamma(c)\Gamma(c-a-b)}{\Gamma(c-a)\Gamma(c-b)} {}_2F_1(a,b;a+b-c+1;1-z) \\
+&\ (1-z)^{c-a-b} \frac{\Gamma(c)\Gamma(a+b-c)}{\Gamma(a)\Gamma(b)} {}_2F_1(c-a,c-b;c-a-b+1;1-z) \ .
\end{split}
\end{equation}

\subsubsection*{Jacobi polynomials}
The Jacobi polynomials are defined in terms of the hypergeometric function as
\begin{equation}\label{JPoly}
P_n^{(l_1,l_2)}(x) = {l_1 + n \choose n } {}_2 F_1\left(-n,l_1+l_2+n+1;l_1+1;\tfrac{1}{2}(1-x)\right) \ .
\end{equation}
They are orthogonal in the interval $[-1,1]$ with respect to the weight $(1-x)^{l_1}(1+x)^{l_2}$:
\begin{equation}
\begin{split}
\int_{-1}^1 (1-x)^{l_1}(1+x)^{l_2}  P^{(l_1,l_2)}_m(x) P^{(l_1,l_2)}_n(x) dx &= \\
\frac{2^{l_1+l_2+1}}{2n+l_1+l_2+1}&\frac{\Gamma(n+l_1+2)\Gamma(n+l_2+1)}{\Gamma(n+l_1+l_2+1)n!} \delta_{mn} \ .
\end{split}
\end{equation}

\bibliography{justabib,at}
\bibliographystyle{justabib}

\end{document}